\documentclass[11pt,twoside]{article}

\usepackage{asp2006}
\usepackage{graphicx}
\begin{document}
\title{Testing stellar cusp formation theories with observations of the Milky Way nuclear star cluster}   
\author{Tuan Do\altaffilmark{1}, Andrea M. Ghez\altaffilmark{1}, Mark R. Morris\altaffilmark{1}, Jessica R. Lu\altaffilmark{2}, Keith Matthews\altaffilmark{2}, Sylvana Yelda\altaffilmark{1}, Shelley Wright\altaffilmark{3}, James Larkin\altaffilmark{1}}   
\altaffiltext{1}{University of California, Los Angeles, CA 90095-1547}    
\altaffiltext{2}{California Institute of Technology, Pasadena, CA 91135}
\altaffiltext{3}{University of California, Berkeley, CA 94720}

\begin{abstract} 
We report on the structure of the nuclear star cluster in the innermost 0.16 pc of the Galaxy as measured by the number density profile of late-type giants. Using laser guide star adaptive optics in conjunction with the integral field spectrograph, OSIRIS, at the Keck II telescope, we are able to differentiate between the older, late-type ($\sim$ 1 Gyr) stars, which are presumed to be dynamically relaxed, and the unrelaxed young ($\sim$ 6 Myr) population. This distinction is crucial for testing models of stellar cusp formation in the vicinity of a black hole, as the models assume that the cusp stars are in dynamical equilibrium in the black hole potential. In the survey region, we classified 77 stars as early-type and 79 stars as late-type. We find that contamination from young stars is significant, with more than twice as many young stars as old stars in our sensitivity range (K$^\prime < 15.5$) within the central arcsecond.  Based on the late-type stars alone, the surface stellar number density profile, $\Sigma(R) \propto R^{-\Gamma}$, is flat, with $\Gamma = -0.26\pm0.24$. Monte Carlo simulations of the possible de-projected volume density profile, n(r) $\propto r^{-\gamma}$, show that $\gamma$ is less than 1.0 at the 99.7 \% confidence level. These results are consistent with the nuclear star cluster having no cusp, with a core profile that is significantly flatter than predicted by most cusp formation theories, and even allows for the presence of a central hole in the stellar distribution. Here, we also review the methods for further constraining the true three-dimensional radial profile using kinematic measurements. Precise acceleration measurements in the plane of the sky as well as along the line of sight has the potential to directly measure the density profile to establish whether there is a ``hole'' in the distribution of late-type stars in the inner 0.1 pc. 

\end{abstract}



\section{Introduction}

Over 30 years ago, theoretical work suggested that the steady state distribution of stars may be significantly different for clusters with a massive black hole at the center than those without \citep[e.g.][]{1976ApJ...209..214B,1978ApJ...226.1087C,1980ApJ...242.1232Y}.  Stars with orbits that bring them within the tidal radius of the black hole are destroyed and their energy is transferred to the stellar cluster. In the steady state, this energy input must be balanced by the contraction of the cluster core, which appears as a steeply rising radial profile in the number density of stars toward the cluster center. This radial profile is usually characterized by a power law of the form $n(r) \propto r^{-\gamma}$, with a power law slope, $\gamma$, that is steeper than that of a flat isothermal core. For a single-mass stellar cluster, \citet{1976ApJ...209..214B} determined the dynamically relaxed cusp will have $\gamma = 7/4$. The presence of such a steep core profile, or cusp, is important observationally because it may represent a simple test for black holes in stellar systems where dynamical mass estimates are difficult, such as in the cores of galaxies. The stellar cusp is also important theoretically as it is a probable source of fuel for the growth of supermassive black holes and its presence is often assumed in simulations of stellar clusters having a central black hole \citep[e.g.,][and references therein]{2005gbha.conf..221M}.

Theoretical work has progressed from the early simulations by \citet{1976ApJ...209..214B} to include many complicating effects, including multiple masses, mass segregation, and stellar collisions, on the density profile of stellar clusters in the presence of a supermassive black hole \citep[e.g.,][]{1977ApJ...216..883B,1991ApJ...370...60M,2009ApJ...697.1861A}. These theories predict cusp slopes within a black hole's gravitational sphere of influence ranging from $ 7/4 \geq \gamma \geq 7/3$ for a multiple mass population to as shallow as $\gamma = 1/2$ for a collisionally dominated cluster core. An important assumption of all cusp formation models is that the stellar cluster be dynamically relaxed. Without this assumption, the stellar distribution would show traces of the cluster origin in addition to the influence of the black hole.

The Galactic center is an ideal place to test these theories of cusp formation as it contains the nearest example of a supermassive black hole \citep[Sgr A*, with a mass M$_{\bullet} = 4.1\pm0.6\times10^{6}$ M$_{\odot}$,][]{2008ApJ...689.1044G,2009ApJ...692.1075G}. Located at a distance of only 8 kpc, the radius of the sphere of influence of the Galactic center black hole ($\sim1$ pc) has an angular scale of $\sim 25\arcsec$ in the plane of the sky, two orders of magnitude larger than any other supermassive black hole.  In order to plausibly test stellar cusp formation theories, the stellar population used to trace the cusp profile must be older than the relaxation time, which is of order 1 Gyr within the sphere of influence of Sgr A* \citep[e.g.,][]{2006ApJ...645.1152H}. In the Galactic center, late-type red giants (K to M) are the most promising tracers of the stellar distribution since they are $> 1$ Gyr old, bright in the NIR, and abundant. On parsec size scales, these stars dominate the flux, so early seeing limited observations of the surface brightness profile in the NIR at the Galactic center were successfully used to show that the structure of the nuclear star cluster at large scales has a density power law of $\gamma \approx 2$ \citep{1968ApJ...151..145B}. Integrated light spectroscopy of the CO absorption band-head at 2.4 $\micron$ (which is dominated by red giants) also confirmed this slope down to within $\sim 0.5$ pc from the black hole \citep{1989ApJ...338..824M,1996ApJ...456..194H}. These measurements, as well as integrated light spectroscopy from \citet{2000ApJ...533L..49F} of the inner 0\arcsec.3, found a lack of CO absorption in the inner $\sim 0.5$ pc of the Galaxy. It was unclear at the time whether this represented a change in the stellar population with fewer late-type stars in the inner region or a change in the stellar density profile. Because the integrated light spectrum is biased toward the brightest stars, contamination from a few young Wolf-Rayet (WR) stars (age $< 6$ Myr), such as the IRS 16 sources identified in the early 1990s located between 1--2$\arcsec$ from the black hole \citep[e.g.,][]{1990MNRAS.244..706A,1991ApJ...382L..19K}, can significantly impact measurements of the underlying stellar density.

Here, we report on an update of our high angular resolution spectroscopic survey of the central 0.15 parsec of the Galaxy using laser guide star adaptive optics with the OSIRIS IFU at the Keck II telescope. The survey reaches a completeness of 40\% at $K^{\prime} \approx 15.5$, two magnitudes fainter than previously reported spectroscopic surveys of late-type stars in this region. We find that the surface number density of late-type stars is significantly flatter than the range of power laws predicted by \citet{1977ApJ...216..883B} (Section \ref{sec:results}). This measurement rules out all values of $\gamma > 1.0$ at a confidence level of 99.7\%. In Section \ref{sec:discussion}, we discuss possible dynamical effects that may lead to the flat observed slope and implications for future measurements of the stellar cusp at the Galactic center. 

\section{Observations}

Near-IR spectra of the central 4$\arcsec$ of the Galaxy were obtained between 2006 and 2009 using the OSIRIS integral field spectrograph \citep{2006NewAR..50..362L} in conjunction with the laser guide star adaptive optics (LGS AO) system on the Keck II telescope \citep{2006PASP..118..297W, 2006PASP..118..310V}. Details of the observations are given in \citep{2009ApJ...703.1323D}. In this paper, we include two additional fields not reported in \citep{2009ApJ...703.1323D}, the northwest and southwest field, extending complete azimuthal coverage in the survey out to $\sim3\arcsec$ from Sgr A* (see Figure \ref{fig:mosaic_fov}). The two additional $2\arcsec\times3\arcsec$ fields are centered at distance of $-1.99\arcsec, 2.4\arcsec$ (northwest) and $-2.99\arcsec,-1.12\arcsec$ (southwest) from Sgr A* (obtained in 2009 May). 

\section{Results}

\begin{figure}[t]
\centering
\includegraphics[width=3.25in,angle=90]{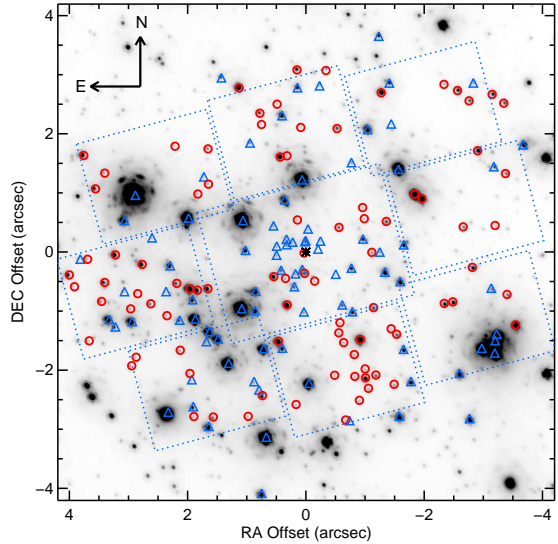}
\caption{The currently surveyed region overlaid on a K$^\prime$ (2.2 \micron) image of the Galactic center taken in 2007. Sgr A* is marked at the center with a *. Spectroscopically identified early (blue triangles) and late-type (red circles) stars are marked. Each field is enclosed by dotted lines. Some spectral identifications are outside of the marked lines because they were found at the edge of the dithers.}
\label{fig:mosaic_fov}
\end{figure}
\label{sec:results}
With spectroscopic identification of the stars brighter than $K^\prime \sim 15.5$, we are able to separate the presumably dynamically relaxed old stars from the unrelaxed young population. The stellar number density profile is constructed in radial bins of 0.\arcsec5 over the area that was sampled in this survey out to a radius of $\sim 3\arcsec$, with error bars scaling as $\sqrt{N}$, where $N$ is the number of stars in each bin. It is clear that, while the number density of early-type stars increases quite rapidly toward the central arcsecond, the late-type stars have a very flat distribution (Figure \ref{fig:mosaic_fov}). Outside the central arcsecond, the projected number density of early-type stars drops to about half that of the late-type stars.

\begin{figure}
\centering
\includegraphics[width=3.5in]{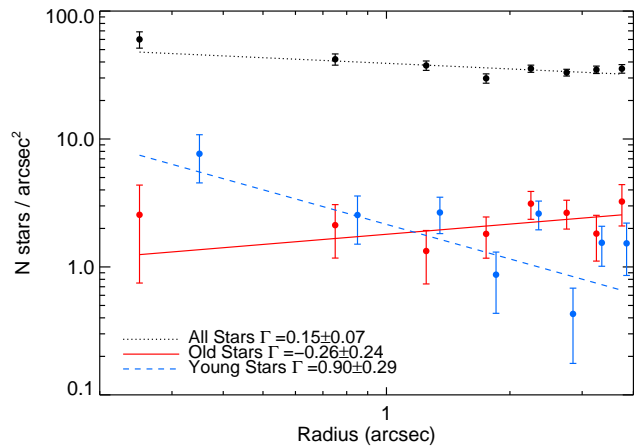}
\caption{Plot of the surface number density as a function of projected distance from Sgr A* in the plane of the sky for different populations: old (late-type, red), young (early-type, blue), and total number counts from $K^\prime$ imaging. These radial profiles have been corrected for completeness and extinction using the method detailed in \citep{2009ApJ...703.1323D}.}
\label{fig:radial_corr}
\end{figure}

Potential systematic errors to the radial profile measurement may arise from variable extinction and incompleteness between each field. In order to study how these two factors impact our result, the radial profile was computed using stars with extinction corrected magnitudes only over regions that are at least 30\% complete. To do so, the extinction map reported by \citep{2009A&A...502...91S} was used to correct for the variations in extinction between each star by adding $A_K - 3.0$ to each star to bring them to the same canonical $A_K = 3.0$ extinction to the Galactic center. Then, the $K^\prime$ luminosity function was recomputed for each field to determine the $K^\prime$ magnitude bin beyond which the completeness falls to less than 30\%. This separates the fields into three groups in terms of their 30\% completeness : all fields are at least 30\% complete down to $K^\prime = 14.5$; all fields except W, NW, and SW are complete to $K^\prime = 15.5$; the central field is complete to $K^\prime = 16.0$. We calculate the radial profiles in each magnitude bin using only fields that are at least 30\% complete at that magnitude. For all stars with $K^\prime < 14.5$, we use all the fields, while for stars with $14.5 \le K^\prime < 15.5$, the western field was dropped from the measurement; the $15.5 \le K^\prime < 16.0$ magnitude bin was not included because only the central field is complete to at least 30\% at that magnitude. The radial profiles from the $K^\prime < 14.5$ and $14.5 \le K^\prime < 15.5$  magnitude bins were then summed to produce the final measurement of the surface density, with errors added in quadrature (see Figure \ref{fig:radial_corr}). The early-type stars have $\Gamma_{young} = -0.90\pm0.27$, while for the late-type stars, $\Gamma_{old} = -0.26\pm0.24$. Neither of these values differ significantly from the case without extinction and completeness correction. The rest of the analysis will use these values for the measured surface number density profile. 

\section{Discussion}
\label{sec:discussion}
The effects of projection of a three-dimensional cluster onto the plane of the sky must be taken into account for quantitative interpretations of the measured surface density profile. The projection onto the sky plane of a spherically symmetric cluster can be done using the integral:
\begin{equation}
\Sigma(R) = 2 \int_{R}^{\infty}\frac{r n(r)dr}{\sqrt{r^{2}-R^{2}}}
\end{equation}
where $R$ is the distance from the center of the cluster in the plane of the sky, $r$ is the physical radius of the star cluster, $\Sigma(R)$ is the projected surface density, and $n(r)$ is the volume number density. Because this integral diverges for all $\gamma < 1.0$, where $n(r) \propto r^{-\gamma}$, the density in the outer region of the cluster must fall off steeply to accommodate a shallow power law at the center of the cluster. While we do not observe a break in the cluster density profile within our $4\arcsec$ survey, one was observed by \citet{2007A&A...469..125S} and others at scales larger than our survey. In order to account for the outer region of the star cluster, we therefore model the number density profile as a broken power law:
\begin{equation}
n(r) \propto \left\{
	\begin{array}{cc}
	r^{-\gamma_{1}} & r < r_{break}, \\
    r^{-\gamma_{2}} & r \ge r_{break}. 
     \end{array}
     \right.
\end{equation}
Using the measurements from \citet{2007A&A...469..125S}, we set $r_{break} = 8\arcsec$ and $\gamma_{2}=1.8$ (the deprojected values of $R_{break}=6\arcsec$ and $\Gamma = 0.8$, from the surface density measurements using this broken power law model) to constrain the parameters of the outer cluster density profile. We assumed that contamination from young stars is less severe in the region outside of our survey, so using an outer cluster density profile measured from total number counts should introduce only a small bias. For example, the density of bright young stars falls off relatively steeply as $R^{-2}$ \citep[][but see also \citet{2009A&A...499..483B}]{2006ApJ...643.1011P,2009ApJ...690.1463L}. To determine the constraints on the physical density profile in the inner region, we performed Monte Carlo simulations of clusters with $\gamma_{1}$ between -2.0 and 2.0. For each value of $\gamma_{1}$, $10^{4}$ realizations were performed. In each realization, 74 stars (the number of late-type stars used to measure the radial profile) were drawn from the broken power law distribution after correcting for the limited area observed in this survey. The locations of these stars were then binned and fitted the same way as the data. We find that for $-2.0 < \gamma_{1} < 0.5$, the projected inner radial profiles are flat with $\Gamma \approx 0$. Using the relationship between $\gamma_{1}$ and $\Gamma$ as determined from the MC simulations and the observed inner radial profile, we can set an upper limit of 1.0 to the value of $\gamma_{1}$ at the 99.7\% confidence level.

\subsection{Mechanisms for cusp depletion}

Although the slope measurement in this survey cannot constrain whether there is a `hole' in the distribution of late-type stars or just a very shallow power law within the central $4\arcsec$, the inferred slope is significantly flatter than the range of $\gamma_{1}$ between 7/4 and 3/2 predicted by \citet{1977ApJ...216..883B}. In fact, a flat core density profile without a cusp can fit the data equally well.  This result is consistent with similar recent measurements from \citet{2009A&A...499..483B} and \citet{2010ApJ...708..834B}. Below, we discuss several plausible dynamical scenarios that may deplete the number density of the late-type giants observed in this study, as well as possible observational prospects for making further progress.

\subsubsection{Mass segregation}
The true population of dark mass that consists of stellar remnants such as neutron stars or stellar black holes is unknown, but theoretically, there is a strong case for a large population of such remnants due to mass segregation \citep{1993ApJ...408..496M}. As these dark remnants migrate inward, they will scatter the lighter stars outward. Recent simulations by \citet{2009ApJ...697.1861A} showed that a population of $\sim 10$ M$_{\odot}$ stellar remnants can sink into a much steeper density distribution ($2 < \gamma < 11/4$) than the lighter stars ($3/2 < \gamma < 7/2$), if the stellar remnants are relatively rare compared to stars. This results in a very dense cluster of stellar remnants, but the prediction of the stellar cusp from strong mass segregation does not differ substantially from the \citet{1977ApJ...216..883B} values. Thus, mass segregation cannot be the only mechanism responsible for the inferred lack of a cusp in the evolved red giant population. 

\subsubsection{Envelope destruction by stellar collisions}

The M and K type giants in this survey have radii that vary between 20 to 200 R$_{\odot}$, which results in substantial cross sections for collisions with other stars in this high density environment \citep[e.g.][]{1999MNRAS.308..257B,2009MNRAS.393.1016D}. This may lead to a preferential depletion of red giants toward the center, which would result in a biased measurement of the underlying stellar cusp slope, since most of the surviving stars in the cusp would be unobserved main sequence stars with smaller radii. The depletion of giants has been suggested in the past based upon low spatial resolution spectroscopy of the CO band-head at 2.29 $\micron$ \citep{1989ApJ...338..824M} as well as in narrow band imaging \citep{2003ApJ...594..812G,2009A&A...499..483B}. Theoretically, a collisionally dominated cusp can be as shallow as $\gamma = 0.5$, which is within the range of slopes allowed by our upper limits, for an isotropic distribution, (i.e., for a distribution function that depends only on energy), although this result no longer holds in the case of a more general distribution function \citep{1991ApJ...370...60M}. While the collisional destruction of giants can operate efficiently within the region observed in this survey, the effectiveness of this mechanism drops very quickly with distance from the black hole. For example, \citet{2009MNRAS.393.1016D} found that collisional destruction of giants becomes less likely beyond about 0.1 pc from Sgr A*. While this survey covers a region out to $\sim0.16$ pc in projection from Sgr A*, previous observations such as from \citet{2009A&A...499..483B} suggest that the flattening in the late-type giant radial profile extends out even further to $\sim0.24$ pc in projection. If the assumptions about the steep distributions of stellar mass black holes and other stellar remnants in the simulations of \citet{2009MNRAS.393.1016D} are correct, then this would suggest that the collisional destruction of giants is not the dominant mechanism for clearing the cusp of stars. Observational constraints on the effectiveness of this mechanism can be made by comparing the radial density profiles of giants of varying stellar radii. Collisional destruction should preferentially remove stars having larger radii at a given distance from the black hole.

\subsubsection{IMBH or binary black hole}

A more dramatic change to the stellar cusp can be achieved through the infall of an IMBH. \citet{2006MNRAS.372..174B} performed N body simulations of the effects on the central star cluster with an initial cusp slope of $\gamma = 7/4$ due to the infall of IMBHs of mass $3\times10^{3}$ M$_{\odot}$ and $10^{4}$ M$_{\odot}$. They find that the IMBH will deplete the cusp of stars, which will cause the inner portion of the cluster to resemble a core profile after the IMBH has spiraled into the central black hole, and that it would take over $\sim$ 100 Myr after the IMBH infall to replenish the stellar cusp. For the case of a $10^{4}$ M$_{\odot}$ IMBH, the density profile has a flat core, with a core radius of $\sim 0.2$ pc, which is roughly consistent with that observed in this survey and by \citet{2009A&A...499..483B}. 

The existence of a binary massive black hole in the Galactic center can also contribute to the loss of stars in this region. A star passing within the orbit of the companion massive black hole will undergo dynamical interactions with the binary, which can result in the star achieving ejection velocity \citep{2005LRR.....8....8M}. In addition, similar to that of an infalling IMBH, a merger with a massive black hole system more recently than the dynamical relaxation time will result in a system that is still out of equilibrium at this time \citep{2007ApJ...671...53M}. 

Current measurements impose only modest limits on the possibility of an infalling IMBH and a binary black hole. \citet{2008A&A...492..419T} found no obvious kinematic signatures of a recent infall of an IMBH using 3D velocity measurements of the late-type stars, however, a more complete study of the phase space distribution of the stars will be necessary to ascertain the likelihood of an IMBH clearing the cusp of stars (see \citet{2009arXiv0905.4514G} for a more extensive discussion of current constraints on the existence of an IMBH at the Galactic center). Current measurements of the reflex motion of the black hole at radio wavelengths place the limit to the mass of a companion black hole at $< 10^4$ M$_{\odot}$ with semi-major axes $10^{3}$ AU $< a < 10^{5}$ AU \citep{2004ApJ...616..872R}. Precise measurements of the possible reflex motion of the black hole using stellar orbits can provide stronger constraints in the future on the presence of a companion black hole \citep{2008ApJ...689.1044G}.

\subsection{Constraining line of sight distance with kinematic measurements}

The degeneracy in the projection from a three-dimensional cluster density profile onto the plane of the sky must be overcome in order to better constrain the theories for the depletion of the stellar cusp. This degeneracy can be resolved by measurements of the line of sight distance to each star. The star cluster in the Galactic center has the unique advantage of having very precise kinematic measurements, which with some assumptions about the potential, can be used to derive the line of sight distances of the stars, $z$, from the black hole. 

The simplest method of determining the limits to $z$ for each star comes from comparing the total velocity, $V_{tot}$ of the star to the escape velocity at the star's projected distance, $R$ from the black hole. If we assume the major contributor of the potential to be the black hole mass, $M_{BH}$, and that the stars are gravitationally bound, then:
\begin{equation}
z < \sqrt{\frac{2GM_{BH}}{V_{tot}^{2}} - R^{2}}
\end{equation}

\begin{figure}
\centering
\includegraphics[width=3in,angle=90]{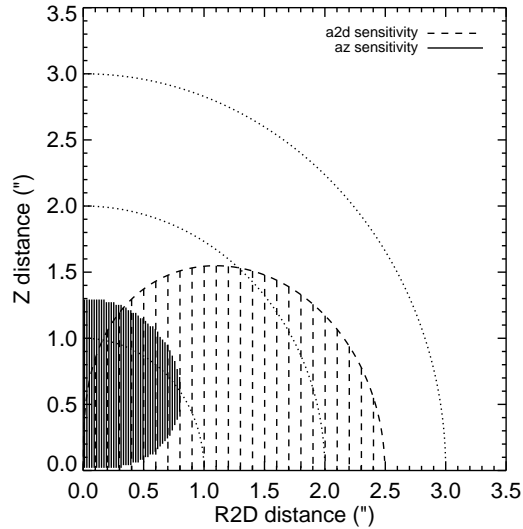}
\caption{The sensitivity to the line of sight distance for stars based on the average acceleration errors along the line of sight ($a_{2D}$) and in the plane of the sky ($a_{z}$) are shown in the shaded regions. The dotted lines show the volume of space encompassed within $1\arcsec, 2\arcsec,$ and $3\arcsec$ from Sgr A*.}
\label{fig:a2d_sensitivity}
\end{figure}

For the stars at the center of the nuclear star cluster, the line of sight distance can be measured directly using accelerations. With the assumption that the black hole mass dominates the potential, the true radial distance, $r$, from the black hole can be simply computed using Newton's law of gravity: 
\begin{equation}
a = \frac{GM_{BH}}{r^{2}}
\end{equation}
where $a$ is the magnitude of the acceleration of the star. This assumes that the acceleration can be measured in both the plane of the sky and along the line of sight. It is useful to decompose these two components, because accelerations in the plane of the sky are measured using astrometry, while the accelerations along the line of sight are measured spectroscopically with different measurement uncertainties. The line of sight distance can be computed from the plane of the sky acceleration, $a_{2D}$, using:
\begin{equation}
\begin{array}{c}
a_{2D} = - \frac{GM_{BH}}{(R^{2}+z^{2})^{3/2}}R \\
z = ((\frac{G M_{BH} R}{a_{2D}})^{2/3} - R^{2})^{1/2}
\end{array}
\end{equation}
With the measurements of $a_{2D}$ alone, $\pm z$ is degenerate (the star can be in front or behind the black hole), but is not of concern for measuring the radial profile of the number density. More problematic however is that as $R_{2D}$ approaches 0, $a_{2D}$ also approaches 0 (as the acceleration vector always points toward the black hole), resulting in less sensitivity to the measurement of $z$ for a given astrometric precision. In contrast, acceleration along the line of sight from spectroscopic measurements are more sensitive for stars close to the black hole in the plane of the sky: 
\begin{equation}
\begin{array}{c}
a_{z} = - \frac{GM_{BH}}{(R^{2}+z^{2})^{3/2}} z
\end{array}
\end{equation}
However, $z$ is degenerate for measurements of $a_{z}$ over a range of $z$. For example, if $a_{z}$ is small, $z$ can be very close to zero, or $z$ can be very large. Limits on $a_{2D}$ are required to break this degeneracy. Figure \ref{fig:a2d_sensitivity} summarizes our sensitivity to $z$ given our average acceleration uncertainties in $a_{z}$ and $a_{2D}$. Currently, we should be able to detect all stars within $1\arcsec$ in volume around Sgr A*. Beyond this volume, we have some sensitivity out to $\sim2.5\arcsec$. With a longer time baseline and lower acceleration uncertainties in measurements of $a_{2D}$, we are able to measure the $z$ distances of stars better than with $a_{z}$ in the survey region. Further details on the measurements and the three dimensional profile of the cluster will be given in Do et al. (in prep). 

\section{Conclusions}
We completed a survey using high angular resolution integral field spectroscopy of the inner $\sim 0.16$ pc of the Galaxy in order to measure the radial profile of the late-type stars in this region. The survey reached a depth of $K^{\prime} < 15.5$ and is able to differentiate between early and late-type stars. We find a late-type stellar surface density power-law exponent $\Gamma = -0.26\pm0.24$, which limits the volume number density profile slope $\gamma$ to be less than 1.0 at the 99.7\% confidence level, and even allows for the presence of a central drop in the density of late-type giants. Given the current measurements, we cannot yet determine whether the distribution of observed giants is representative of the distribution of stellar mass. Being able to infer the underlying dynamically relaxed stellar population will be crucial in order to establish whether the Galactic center is lacking the type of stellar cusp long predicted by theory. Obtaining an unbiased measurement of the stellar distribution is important because such cusps have considerable impact on the growth of massive black holes as well as on the evolution of nuclear star clusters.  Progress in achieving this goal will be possible with improved kinematics and spectral coverage in order to break the degeneracy in the surface number density profile to better establish the three-dimensional distribution of the stellar cluster.   

\acknowledgements 
upport for this work was provided by NSF
grants AST-0406816 and AST-0909218, and the NSF Science
\& Technology Center for AO, managed by UCSC
(AST-9876783).
The infrared data presented herein were obtained at the W. M. Keck Observatory, which is operated as a scientific partnership among the California Institute of Technology, the University of California and the National Aeronautics and Space Administration. The Observatory was made possible by the generous financial support of the W. M. Keck Foundation. The authors wish to recognize and acknowledge the very significant cultural role that the summit of Mauna Kea has always had within the indigenous Hawaiian community. We are most fortunate to have the opportunity to conduct observations from this mountain.

\end{document}